\documentclass[journal]{IEEEtran}

\usepackage[left=0.75in,right=0.75in,top=0.75in,bottom=0.75in]{geometry}

\usepackage{color}
\usepackage{qtree}
\usepackage{times}
\usepackage{epsfig}
\usepackage{graphicx}
\usepackage{epstopdf}
\usepackage{algpseudocode}
\usepackage{algorithm}
\usepackage{amsmath}
\usepackage{amssymb}
\usepackage{amsxtra}
\usepackage{multirow}
\usepackage{mathtools}
\usepackage{cleveref}
\usepackage{url}
\usepackage{cite}
\usepackage{amsthm}
\usepackage{adjustbox}
\usepackage[latin1]{inputenc}
\usepackage{caption}
\usepackage[caption = false]{subfig}
\usepackage{float}

\begin{document}
\title{PhD Forum: Enabling Autonomic IoT for Smart Urban Services}
%Thinking Online and Thinking Ahead: Strategic Decisions for Smart Urban Services}
%Integrating Supply Chain into Risk Analysis of IoT Systems}
%Risk Analysis of Supply Chain Attacks in IoT Systems
%Analyzing Hierarchical Supply Chain Threats in IoT Systems
%Holistic Risk Analysis of Hierarchical Supply Chain Threats in IoT Systems

\author{ \IEEEauthorblockN{\large Muhammad Junaid Farooq} and \IEEEauthorblockN{\large Quanyan Zhu} \\ \IEEEauthorblockA{Department of Electrical \& Computer Engineering, Tandon School of Engineering, \\New York University, Brooklyn, NY 11201, USA,} Emails: \{mjf514, qz494\}@nyu.edu. \vspace{-0.3in}
%{\thanks {\vspace{-0.2cm}\hrule \vspace{0.2cm} \indent This work was made possible by.}}

%\thanks{\vspace{-0.2in}\hrule \vspace{0.2cm}
%Muhammad Junaid Farooq and Quanyan Zhu are with the Department of Electrical \& Computer Engineering, Tandon School of Engineering, New York University, Brooklyn, NY, USA, E-mails: \{mjf514, qz494\}@nyu.edu.}
}

\maketitle

\begin{abstract}
The development of autonomous cyber-physical systems (CPS) and advances towards the fifth generation (5G) of wireless technology is promising to revolutionize many industry verticals such as Healthcare, Transportation, Energy, Retail Services, Building Automation, Education, etc., leading to the realization of the smart city paradigm. The 
%cyber-physical integration of smart devices (sensors and actuators) in such systems, referred to as the
Internet of Things (IoT), enables powerful and unprecedented capabilities for intelligent and autonomous operation. We leverage ideas from Network Science, Optimization \& Decision Theory, Incentive Mechanism Design, and Data Science/Machine Learning to achieve key design goals, in IoT-enabled urban systems, such as efficiency, security \& resilience, and economics.

\end{abstract}

\begin{IEEEkeywords}
Internet of things, cyber-physical systems, mission-critical, network science. \vspace{-0.1in}
\end{IEEEkeywords}

\maketitle

\section{Introduction}

%The IoT comprises of a network of sensors and actuators, which are embedded computers, communicating with each other and to the Internet. 
The cyber-physical integration of devices in the IoT is enabling the development of a myriad of applications and services. 
%However, it is not a standalone system. In fact, the IoT ecosystem is extremely diverse and involves an interplay of many different systems ranging from endpoint devices, communication networks, cloud computing systems, and user devices. 
%Furthermore, these components are often operated and controlled by totally different entities. 
At each layer of interaction between the systems, there are decision problems that arise for achieving various different objectives. Effective decision-making at different fronts is essential to enhance the efficiency, economics, security, and resilience of these systems. This research focuses on developing cyber-physical decision mechanisms. 
%that are well suited for various different levels across the IoT ecosystem. 
For instance, at the device level, it is important to ensure wireless connectivity and communication. If the communication infrastructure is in place, there is a need for spectrum allocation and reservation. However, in case of challenged infrastructure, the connectivity might be achieved using an overlay network of aerial base stations~\cite{uav_tccn}
\cite{uav_globecom}. Decisions have to be made on allocation of spectrum to users and configuration/placement of aerial base stations to provide adequate coverage and connectivity. Once the connectivity is achieved, the networks are used for information dissemination~\cite{massive_iot}. 
%The goal is to design networks that have the desired information propagation and are re-configurable in the case of adversarial attacks. 
Furthermore, there is a need to design security mechanisms against stealthy adversarial threats that may be using the same communication networks to infiltrate and sabotage network operation. The next frontier is the use of cloud computing resources by smart devices. It is important to efficiently allocate and price the computational resources in order to provide a high quality of experience to the users and to generate high revenues for the cloud/fog service provider. Finally, at the application front, there are scenarios where resource provisioning decisions need to be made for service requests that appear randomly in space and time as in instance in the case of smart urban environments.

\begin{figure}[t]
\centering
\includegraphics[width = 3in]{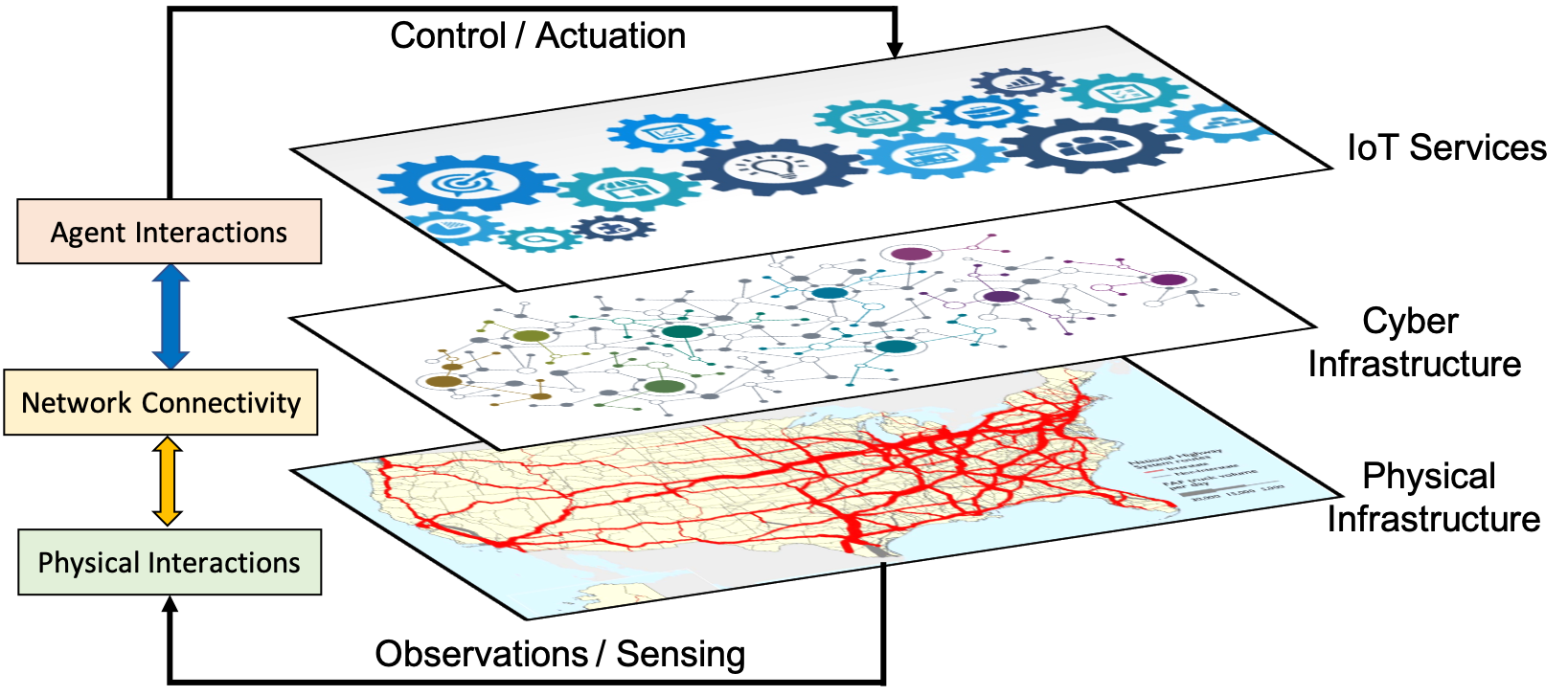}
      \caption{Cross-layer decision-making in the IoT ecosystem.\vspace{-0.1in}}
      \label{pic1}
\end{figure}

\vspace{-0.1in}
\section{Challenges}
Dynamic mechanism design traditionally has been focused on principal-agent type of models with a mechanism designer and participants. In IoT systems, these mechanisms may not always be available. The IoT provides new avenues for designing policies and mechanisms in a dynamic setting at multiple different levels. For instance, at the physical layer, there is a need for connectivity mechanisms while at the higher layers, policy decisions such as allocation and pricing are required. Cyber-Physical coupling needs to be explored in the realm of the IoT which is non-existent in traditional communication and computing systems. Moreover, the IoT ecosystem is inherently a large-scale, complex, and dynamic. Therefore, centralized mechanisms are difficult to implement and are in-feasible. Therefore, more distributed approaches are required.

\vspace{-0.1in}
\section{Contributions}

Autonomous operation of CPS/IoT systems requires an interdisciplinary and cross-layer approach due to the coupling between cyber and physical components. The state and communication at the cyber layer influences the dynamics and control at the physical layer and vice versa. 
%Hence, a cross-layer and distributed approach is crucial for effective design and operation of such systems, particularly for massive systems-of-systems scenarios. 
An high level illustration of the system is provided in Fig.~\ref{pic1}. The following subsections provide details on some of the key thrusts of this research. 

\vspace{-0.1in}
\subsection{Autonomic Networked CPS: From Military to Civilian Applications}

\begin{figure}[t]
\centering
\includegraphics[width = 3in]{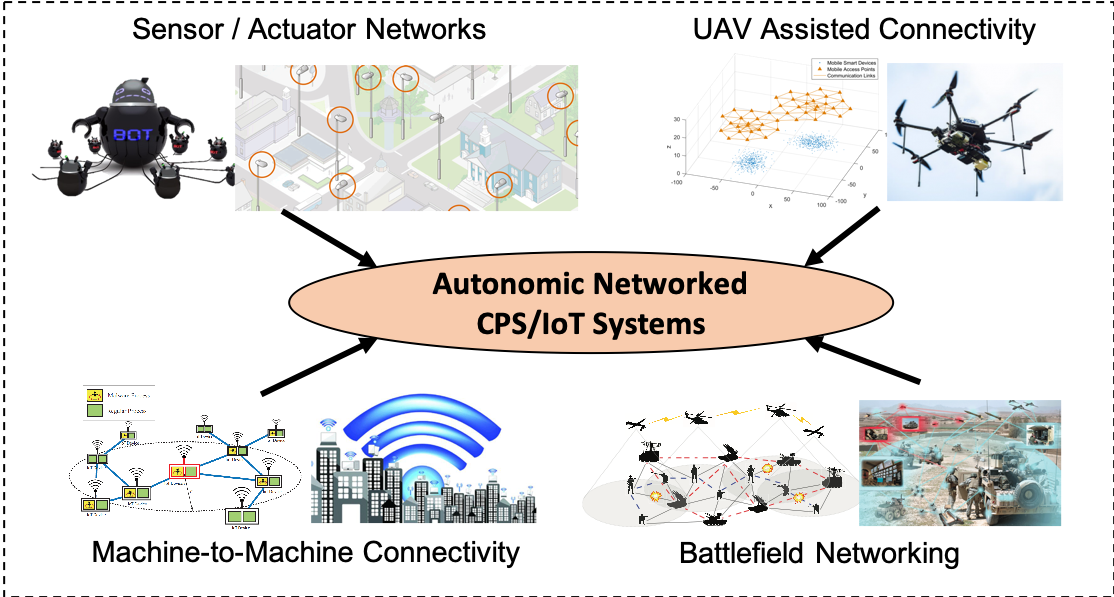}
      \caption{Autonomic CPS/IoT Systems.\vspace{-0.1in}}
      \label{pic2}
\end{figure}

Connected CPS networks are used for dissemination of
data and information for enhanced situational awareness
and decision-making. Some example applications are illustrated in Fig.~\ref{pic2}. 
%Hence, it is imperative to design the networks in a way that ensures that information reliably propagates throughout the network and also ensures that networks have the capability to recover from failures/attacks that may sabotage operation. 
Of particular interest are the military and tactical networks with stringent requirements such as support for extreme
heterogeneity, rapid re-configurability, and mosaic warfare needs. In this respect, one of the key contributions of our research is the development of a secure and re-configurable network design framework suitable for adversarial environments such as the Internet of Battlefield Things (IoBT)~\cite{iobt_twc}\cite{iobt_wiopt}. Networks with enhanced data dissemination capabilities also open doors for malicious activity and malware. Furthermore, malicious entities such as compromised supply chain actors may exploit backdoor channels for stealthy takeover and cause large scale coordinated attacks. To tackle this security risk posed by wireless connectivity proliferation in IoT, we developed a mechanism for inspecting devices based on their wireless neighbourhood that prevents a large scale coordinated attack while causing minimum operational interruption~\cite{tifs_junaid}. 

\subsection{Strategic Resource Provisioning for Mission-Critical IoT Services}

%Limited available resources need to be effectively allocated in an IoT ecosystem at various different levels. 
At the communication layer of the IoT, the spectrum resources needs to be effectively provisioned to applications according to their performance requirements and power limitations~\cite{contract}. Similarly, at the cloud layer, computing nodes and data processing resources need to be allocated and priced strategically to ensure maximum revenue for the cloud service provider. In this regard, we have developed a real-time resource allocation and pricing framework for cloud-enabled IoT systems, where the computational complexity of arriving tasks is evaluated and is accordingly assigned to the available computing bundles~\cite{acc_cloud}. Similarly, for low latency applications, a framework is developed to appropriately select the resources available at one of the fog/edge computing nodes for real-time tasks~\cite{qoe_cloud}. %These works extend the existing works on cloud computing by incorporating the real-time aspect. 
We also extend the allocation and pricing framework to the case where there is uncertainty in the spatial domain in addition to the temporal domain~\cite{spatio-temporal}.

\subsection{Data-Driven Decisions for Urban Service Provisioning}
%The availability of contextually rich data streams in the urban setting along with advances in machine learning and computational capabilities provide an untapped opportunity to analyze and learn human behaviours. It allows us to understand, preferences, incentives, and relationships that act as a basis to develop mechanisms for achieving desired goals and objectives.

\begin{figure}[t]
\centering
\includegraphics[width = 3in]{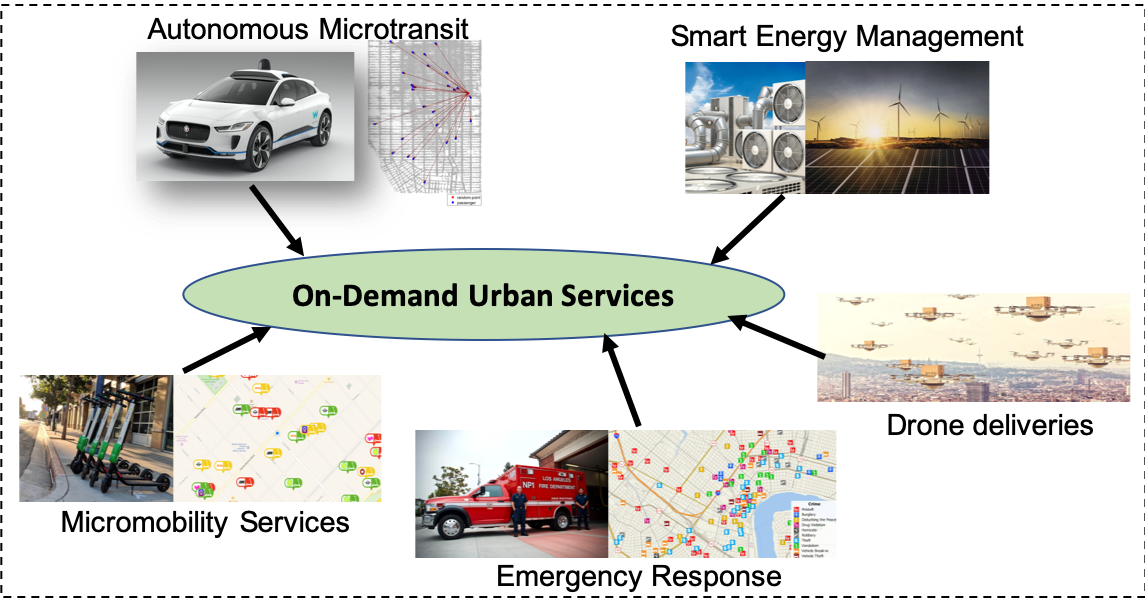}
      \caption{Examples of urban on-demand services.\vspace{-0.1in}}
      \label{pic3}
\end{figure}

IoT-driven urban services are rapidly emerging in almost all industry verticals. Companies like Uber, lyft, Via, etc. have come up with a range of on-demand urban mobility solutions and are moving towards autonomous microtransit solutions. Similarly, there is a strong interest in autonomous vehicle and drone based delivery services. These applications require intelligent decision-making by autonomous agents. Emergency response and first responder tactical systems are key to the safety and resilience of smart cities. The use of sensing capabilities, historical data, and distributed data processing can not only assist early detection and rapid response, but can also help in emergency preparedness. For instance, wildfires are natural disasters that pose a significant threat to the metropolis in terms of life, health, safety, etc. The use of data from sensors reporting temperatures along with weather data can help in determining the timing and location of such incidents. Similarly, timely emergency response is critical for urban safety and quality of life. Some example urban services are depicted in Fig.~\ref{pic3}.

\section{Conclusion}
%Networks are rapidly moving from providing internet connectivity towards service delivery over the IoT. 
%Networks and devices are becoming more and more autonomous to support smart services particularly in the urban context.
%Examples are Autonomous vehicle based ride sharing. Decisions need to made in real-time based on goals such as security, resilience, efficiency, economic-s, etc.
%This research is focused on developing new methods to think online and think ahead to solve these problems. Design mechanisms that allow for real-time and anticipative decision-making. 
%The eventual goal is to move from data to decisions directly without using intermediate models.
This research is an attempt to lay the theoretical foundations of decision science in IoT network design and operation. It leverages tools and theories from various different systems sciences such as mathematical epidemiology, spatial point processes, stochastic processes, optimal control theory, and optimization to tackle these challenging problems. It addresses the challenges and problems at various different levels across the IoT stack. 
%However, it has only scratched the surface of the immense possibilities of dynamic mechanism design in wireless and IoT systems \& networks. 
We hope that this work will form the basis for the development of a comprehensive science for decision mechanisms in the IoT networks domain.

%Towards Service Delivery using networks

\vspace{-0.1in}
\bibliographystyle{IEEEtran}
\bibliography{references}

\end{document}